\documentclass[doublecol]{epl2} 
\usepackage{subcaption}
\usepackage[colorlinks,linkcolor=blue,citecolor=blue,urlcolor=blue,bookmarks=false,hypertexnames=true]{hyperref}
\usepackage{float}
\usepackage{comment}
\usepackage{mathtools}
\usepackage{graphicx}
\usepackage[export]{adjustbox}
\usepackage{dcolumn}
\usepackage{bm}
\title{Probing Majorana Bound States via Thermoelectric Transport}
\author{Colin Benjamin\thanks{\href{mailto:colin.nano@gmail.com}{colin.nano@gmail.com}}\inst{1,2} and Ritesh Das\inst{1,2}}
\shortauthor{Colin Benjamin \etal}
\institute{ School of Physical Sciences, National Institute of Science Education \& Research, Jatni-752050,\ India                    \inst{1} \\ Homi Bhabha National Institute, Training School Complex, Anushaktinagar, Mumbai 400094, India  \inst{2}}
\abstract{We propose a set of thermoelectric experiments based on Aharonov-Bohm interferometry to probe Majorana bound states (MBS), which are generated in 2D topological insulators (TI) in the presence of superconducting and ferromagnetic correlations via the proximity effect. The existence and nature (coupled or uncoupled) of these MBS can be determined by studying the charge and heat transport, specifically, the behavior of various thermoelectric coefficients like the Seebeck coefficient, Peltier coefficient, thermal conductance, and violations of Wiedemann-Franz law as a function of the Fermi energy and Aharonov-Bohm flux piercing the TI ring with the embedded MBS.}
\begin{document}
\maketitle
\section{Introduction}
Current research in quantum computation is concerned with finding materials that can be used in the design of quantum computers. The main obstacle to the fabrication of quantum computers is decoherence and the generation of errors. Quantum computers with topological characteristics can effectively neutralize the effect of decoherence and errors \cite{sankar}. Majorana bound states (MBS) are known to occur in topological insulators \cite{KITAEV20062, MBS1,mbs2, nilssonMBS} in conjunction with superconductors \cite{ivanov, bolech} and ferromagnets, and have the unique property of being their anti-particle. One can encode information in these MBS, thus, protecting them from errors \cite{KITAEV20062}. Methods to probe MBS have been proposed by one of us\cite{colin} using Aharonov-Bohm interferometry that relies on the symmetry of the non-local conductance to Aharonov-Bohm flux. {Further, interference of a different sort via Andreev double dot interferometer \cite{aksenov} has also been used to detect MBS. Similarly, our original idea of detecting MBS by looking at the non-local  Aharonov-Bohm conductance\cite{colin} has been extended to look at Coulomb blockade regime in\cite{fuKaneMBS} or the periodicity of the oscillations in\cite{mesaros, hell}, and to experiments in\cite{nichele} to detect MBS.} \par

In this work, we aim to probe MBS in topological insulators that contain a superconducting and ferromagnetic interface (STIM interface) \cite{nilssonMBS} by studying the symmetry of various thermoelectric coefficients. We see that thermoelectric coefficients like Seebeck, Peltier coefficients, and the thermal conductance \cite{sothmannog, sothmannoptimal, Whitney, strainedheat,strainedrefg} are symmetric or asymmetric to the Aharonov-Bohm flux, Fermi energy and can indicate existence and nature of MBS. Several methods of probing MBS via thermoelectric transport have been put forward before \cite{ramosweideman, thermodonebefore, Leijnse_2014, thermdetect}. The methods proposed in \cite{ramosweideman, thermodonebefore, Leijnse_2014, thermdetect} rely on the magnitude of thermoelectric coefficients like the Seebeck coefficient to probe the presence and nature of MBS. On the other hand, we rely on the symmetry of the thermoelectric coefficients to the Aharonov-Bohm flux and the Fermi energy to probe the absence or presence of MBS as well as their nature (coupled/individual). Additionally, Ref. \cite{ramosweideman} relies on the violation of Wiedemann-Franz law (WF law) \cite{Mermin}. In this paper, we, too, employ the violation of WF law to probe the presence of MBS. An upshot of this work is that none of the previous works that aim to probe MBS via thermal coefficients use Aharonov-Bohm flux as a parameter.
\par
The rest of the paper is organized as follows: the next section discusses thermoelectric transport in mesoscopic systems with two terminals. We use the Onsager matrix that relates heat and charge current to the thermodynamic forces (voltage and temperature bias) and derive the expressions for different thermoelectric coefficients. Next, we elaborate on the proposed model used as a probe for MBS and derive the scattering amplitudes and the transmission probability that will be used to calculate the thermoelectric coefficients. Then we show the variation of thermoelectric coefficients such as Seebeck, Peltier, and the thermal conductance versus the Aharonov-Bohm flux and Fermi energy, both in the absence and presence of MBS. When an Aharonov-Bohm flux is introduced, the thermoelectric coefficients are asymmetric to Fermi energy only when MBS are present and coupled. In the absence of MBS, the Seebeck and Peltier coefficients vanish, while the thermal conductance remains constant. We also plot the thermoelectric coefficients versus the coupling energy of the MBS. We see that in the absence of magnetic flux, the thermoelectric coefficients are symmetric with respect to the coupling between the MBS, however, in the presence of a magnetic flux, the thermoelectric coefficients become asymmetric. Next, we study violation of the Wiedemann-Franz law. We show that WF law is only violated when MBS are present, regardless of the coupling. When MBS are absent, there is no violation of WF law. Further, we see that the WF ratio (i.e., the ratio of electrical conductance to thermal conductance) behaves similarly to other thermoelectric coefficients in terms of symmetry versus Aharonov-Bohm flux, as well as Fermi energy and thus, can also be used to probe the existence and nature of MBS. We summarize the outcomes of our study for investigating MBS in Table 1. We end with a conclusion, summarizing our results.
\section{Theory of Thermoelectric transport in mesoscopic systems}
Mesoscopic charge and heat transport in two terminal setups can be effectively described by scattering matrix theory \cite{backscatter, Butcher_1990,sothmannog}. We denote current by the vector $I = (I_{c}, I_{q}),$ where $I_{c}$ denotes the charge and $I_{q}$ the heat current, and the thermodynamic force vector is defined as $F = (V, \Delta T)$ with $V$ being voltage bias while $\Delta T$ is the temperature difference across the two terminals. The Onsager matrix relating current ($I$) with the thermodynamic forces ($F$) is given by $I = LF$, where \cite{sothmannoptimal, Butcher_1990,strainedheat},
\begin{subequations}
\begin{equation}
\resizebox{1\hsize}{!}{%
$L = \begin{pmatrix}
L_{cV} & L_{cT}\\
L_{qV} & L_{qT}
\end{pmatrix}
= \dfrac{1}{h}\int^{\infty}_{-\infty}dE\overline{T}(E, E_{F})
\xi(E, E_{F})M(E, E_{F}),$
}
\end{equation}
\begin{equation}
\text{with }M(E, E_{F}) = G_{0}
\begin{pmatrix}
1 & (E - E_{F})/eT\\
(E-E_{F})/e & (E-E_{F})^{2}/e^{2}T
\end{pmatrix},
\end{equation}
\begin{equation}
\text{and, } \xi(E, E_{F}) =\dfrac{-\partial f(E, E_{F})}{\partial E},
\end{equation}
where $f(E, E_{F})= \dfrac{1}{1 + e^{(E-E_{F})/k_{B}T}}$ is the Fermi function.
\end{subequations}
In Eq. (1) $T$ is temperature, $k_{B}$ is Boltzmann constant, $G_{0} = (e^{2}/\hbar)$, while $L_{cV} = \sigma $ defines the electrical conductance for the setup, $E$ is incident electron energy, $E_{F}$ is Fermi energy, $\overline{T}$ the transmission probability that includes contribution from both electrons and holes with $h$ being Planck's constant. The coefficients $L_{cV}$ and $L_{qT}$ in Eq.~(1) are related to electrical and thermal conductance, while off-diagonal elements $L_{cT}$ and $L_{qV}$ are related to Seebeck and Peltier Coefficients.\\
From Eq. (1) relating the charge and heat currents with thermodynamic forces within linear irreversible thermodynamics, we can write \cite{beneticasati}:
\begin{equation}
\begin{split}
I_{c} = L_{cV}V + L_{cT}\Delta T,\\
\text{and, } I_{q} = L_{qV}V + L_{qT}\Delta T.
\end{split}
\end{equation}
The Seebeck coefficient is defined as the voltage bias $V$ generated across the terminals when a unit temperature difference $\Delta T$ is applied in the absence of charge current. The Peltier coefficient, on the other hand, is the ratio of the heat current to the charge current across the system in the absence of any temperature difference. From Eqs. (1) and (2), we can write the Seebeck and Peltier coefficients as \cite{beneticasati, strainedheat},
\begin{equation}
S = \dfrac{-L_{cT}}{L_{cV}} \text{, and } P = \dfrac{L_{qV}}{L_{cV}}.\\
\end{equation}
The thermal conductance is the amount of heat current generated due to a unit temperature bias in the absence of charge current. Thermal conductance $\kappa$ is given as,
\begin{equation}
\kappa = \dfrac{I_{q}}{\Delta T}|_{I_{c} = 0} =\dfrac{L_{cV}L_{qT} - L_{cT}L_{qV}}{L_{cV}}.
\end{equation}
Using transmission probability ($\overline{T}$) in Eq. (1), we can determine the Onsager coefficients, $L_{cV}, L_{qV}, L_{cT}, L_{qT}$, and rest of the thermoelectric coefficients.\\
In the next section, we discuss the Aharonov-Bohm interferometer used in our proposed experiment to probe MBS and study the edge mode transport in the interferometer, which determines the transmission probability.
\section{Model}
\subsection{Hamiltonian}
Fig. 1 (a) shows the 2D TI used to generate and probe MBS \cite{nilssonMBS, colin, mesaros}. Our setup is an Aharonov-Bohm interferometer (ABI) \cite{ab} made of 2D TI (e.g. HgTe or CdTe quantum well) \cite{hgtecdte, hgte}. Spin-orbit coupling generates protected 1D edge modes in the 2D TI. An Aharonov-Bohm flux $\Phi$ pierces the ring, as shown in Fig. 1 (a). The ring is connected to two leads via couplers (shown as trapeziums) which in turn are connected to two terminals at temperatures $T_{1}, T_{2}$ and voltages $V_{1}, V_{2}$ respectively. Schrodinger's equation for the electron and hole edge modes in the upper and lower arms of the ring is \cite{colin}:
\begin{equation}
[v_{F} p \tau_{z} \sigma_{z} + (- E_{F} + eA/(\hbar c))\tau_{z}]\psi = E\psi\, ,
\end{equation}
with $p = -i \hbar \partial/\partial x $ being the momentum operator, $E_{F}$ being the Fermi energy, $E$ being the incident electron energy, $v_{F}$ being Fermi velocity, and $A$ being the magnetic vector potential. $\psi$ is a four component spinor given by $\psi = (\psi_{e \uparrow}, \psi_{e \downarrow},\psi_{h \downarrow},\psi_{h \uparrow})^{T}$. The $\tau$ matrices are Pauli matrices that cause mixing in the electron and the hole blocks of Hamiltonian.
MBS occurs at the interface of the Superconducting (S) and Ferromagnetic (F) layers in the upper half of the ring, as shown in Fig. 1 (c). The MBS are denoted as orange ellipses. The superconductor and ferromagnet present in the upper arm of the TI induce superconducting and ferromagnetic correlations in the sample via the proximity effect. Topological edge modes occur and circulate along the edges of the 2D TI and interact with the couplers. MBS appears at the place where the edge modes in TI intersect with the superconducting-ferromagnetic interface \cite{FuKaneMBSog, fuKaneMBS}.
\begin{figure}
\center
\includegraphics[width=0.5\textwidth]{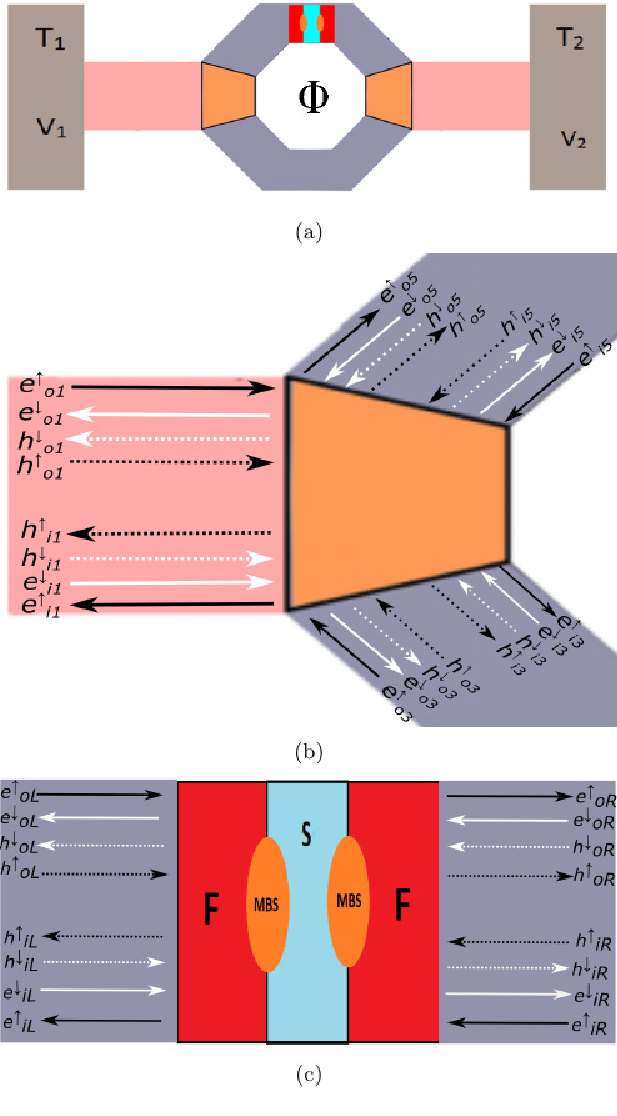}
\caption{(a) The MBS-AB interferometer. A 2D TI fashioned as a ring is connected via two couplers (shown as trapeziums in orange) to leads that are connected to reservoirs at temperatures $T_{1}, T_{2}$ and voltages $V_{1}, V_{2}$. An Aharonov-Bohm flux pierces the ring. (b) Scattering of the edge modes at the left coupler (one can similarly construct the right coupler by following the description in section III B). (c) MBS scatterer. MBS (shown as orange ellipses) occurs at the interface of the Ferromagnetic (shown in red) and superconducting (shown in blue) regions. In (b) and (c), the dashed lines represent the hole edge modes, and the solid lines represent the electron edge modes. The black lines denote spin-up edge modes, and the white lines denote spin-down edge modes. The outer edge modes are shown as double headed arrows while the inner edge modes are shown as single headed arrows.}
\end{figure}
The Hamiltonian for the MBS is \cite{colin, nilssonMBS},
\begin{equation}
H_{M} = -\sigma_{y}E_{M}\, ,
\end{equation}
where $E_{M}$ denotes the coupling between the individual MBS. The edge modes interact with the MBS (as shown in Fig. 1(c)) and the couplers (as shown in Fig. 1(b)). In the next subsection, we discuss the scattering of the edge modes in the system and calculate the transmission probability $\overline{T}$.
\subsection{Transport in the system via edge modes}
To understand scattering in our system, we first describe a related setup, a simple quantum Hall conductor with an Aharonov-Bohm flux. A localized state sensitive to flux develops around the hole in the ring, while edge modes that are insensitive to flux develop on the leads. To couple the edge modes in the ring to the edge modes in the leads, we introduce two couplers (see Fig. 1 (b)) in the system that serves to couple the inner and outer edge modes by inducing inter-edge scattering (See Ref. \cite{colin}). In the case of a topological insulator (TI), the edge modes occur in pairs and have opposite spins. The couplers induce backscattering in all the edge modes in the ring. In addition to backscattering, the MBS mixes the electron edge modes with the hole edge modes via Andreev reflection \cite{nilssonMBS}.
\begin{figure}
\center
\includegraphics[width=0.5\textwidth]{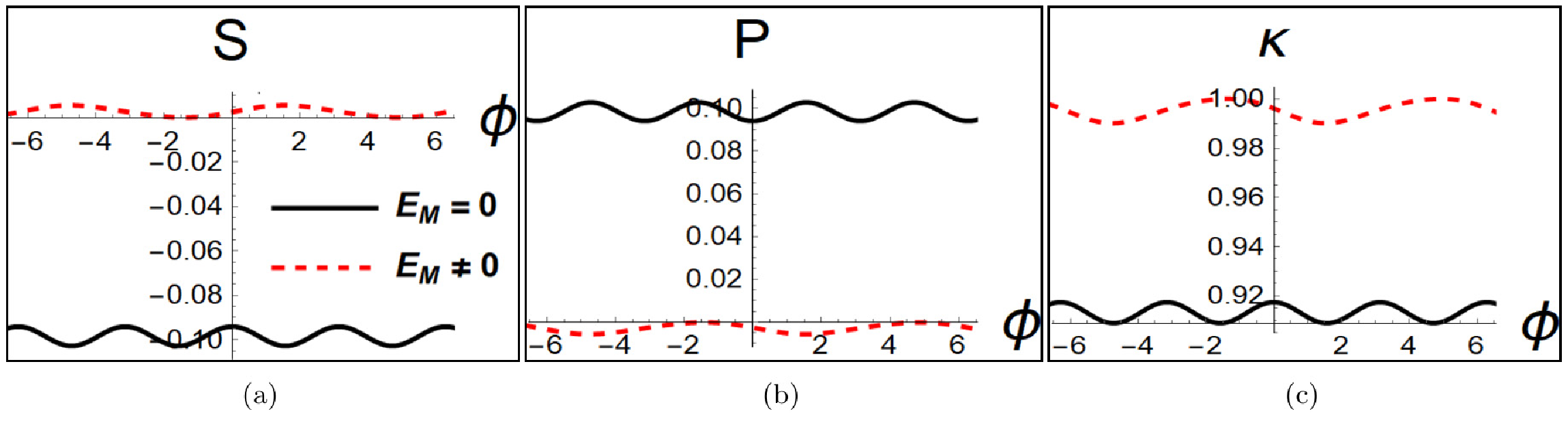}
\caption{(a) Seebeck coefficient $S$ in units of $k_{B}/eT$, (b) Peltier coefficient $P$ in units of $k_{B}/e$ and (c) thermal conductance $\kappa$ in units of $\kappa_{0} = \pi^{2} k_{B}^{2} T/3h$ vs. {the dimensionless Aharonov-Bohm flux $\phi = \Phi / \Phi_{0}$, where $\Phi$ is the Aharonov-Bohm flux, and $\Phi_{0}$ is the flux quantum}, $E_{F} = 3 \mu eV$, $T = 115mK$, for $E_{M} = 0$ (individual MBS) and $E_{M} = 10 \mu eV$ (coupled MBS).}
\end{figure}

A total of four outer edge modes and four inner edge modes occur in the interferometer. However, in the absence of spin-flip scattering, the scattering between edge modes of opposite spin is absent, allowing us to divide the edge modes into two sets. The first set with spin-up electron and spin-up hole edge modes only (shown in black in Fig. 1) and a second set with counterpropagating spin-down electron and spin-down hole edge modes only (shown in white in Fig. 1). The couplers couple outer and inner edge modes. A $6 \times 6$ matrix describes scattering by the couplers. The MBS scatterer couples the incoming spin-up electron and spin-up hole edge modes to outgoing spin-up electron and spin-up hole edge modes. For the first set, incoming edge modes are denoted as $I_{m1} = (e^{\uparrow}_{oL}, e^{\uparrow}_{iR}, h^{\uparrow}_{oL}, h^{\uparrow}_{iR})$ while outgoing edge modes are denoted as $O_{m1} = (e^{\uparrow}_{oR}, e^{\uparrow}_{iL}, h^{\uparrow}_{oR}, h^{\uparrow}_{iL})$ (see Fig. 1 (c)). The propagation and scattering in the second set is the exact mirror image of the scattering of edge modes in the first set. The incoming edge modes for second set are denoted as, $I_{m2} = (e^{\downarrow}_{oR}, e^{\downarrow}_{iL}, h^{\downarrow}_{oR}, h^{\downarrow}_{iL})$ while outgoing edge modes are denoted as $O_{m2} = (e^{\downarrow}_{oL}, e^{\downarrow}_{iR}, h^{\downarrow}_{oL}, h^{\downarrow}_{iR})$. The scattering can be described by a $4 \times 4$ matrix $S_{Maj}$ such that, $I_{mi} = S_{Maj} O_{mi}, i \in \{1,2\}$. As the scattering in both sets mirrors each other, the total transmission probability will be the same for both cases due to symmetry between the edge modes. Thus, we can calculate the transmission probability for one of the sets and then double it to get the total transmission probability.\\
We will now describe this scattering process of edge modes in the first set mathematically. {The $4\times 4 $ scattering matrix for the MBS scatterer is given by \cite{colin},
\begin{subequations}
\begin{equation}
S_{Maj}=\left(\begin{array}{cccc} s_{11}^{ee}&s_{11}^{eh}&s_{12}^{ee}&s_{11}^{eh}\\
	s_{11}^{he}&s_{11}^{hh}&s_{12}^{he}&s_{12}^{hh}\\
	s_{21}^{ee}&s_{21}^{eh}&s_{22}^{ee}&s_{12}^{eh}\\
	s_{21}^{he}&s_{21}^{hh}&s_{22}^{he}&s_{22}^{hh}\end{array}\right)
\end{equation}
where,
\begin{eqnarray}
s_{11}^{ee} =s_{11}^{hh}=1+s_{11}^{eh}=1+s_{11}^{he}=1+i{\Gamma_{1}(E + i\Gamma_{2})/z}, \nonumber\\ s_{22}^{ee} = s_{22}^{hh} =1+s_{22}^{eh}=1+s_{22}^{he}= 1+i{\Gamma_{2}(E + i\Gamma_{1})/z},\nonumber\\
s_{21}^{ee}=-s_{12}^{ee}=s_{21}^{eh}=-s_{12}^{eh}={E_{M}\sqrt{\Gamma_{1}\Gamma_{2}}/z},\\s_{21}^{hh}=-s_{12}^{hh}=s_{21}^{he}=-s_{12}^{he}={E_{M}\sqrt{\Gamma_{1}\Gamma_{2}}/z},\nonumber\\
 z=E_{M}^{2}-(E + i\Gamma_{1})(E + i\Gamma_{2})\nonumber,
\end{eqnarray}
\end{subequations}
and $\Gamma_{1}$ and $\Gamma_{2}$ are the strengths of the couplers coupling the MBS to the left and right arms of the upper ring. Time reversal symmetry is broken in the setup because of two parameters. The first, a non-zero Aharonov-Bohm flux,  and the second coupled Majorana bound states indicated via $E_{M} \neq 0$. The reason for preservation of time-reversal symmetry for $\phi=0$, is that in the S-Matrix which is derived from the Hamiltonian at $\phi=0$, does indeed preserve time reversal symmetry and for this $\phi$ value there is no Aharonov-Bohm flux.  Further, for the case of coupled Majorana bound states, one can see this from the S-Matrix, $S_{Maj}$ defined in Eq.~7, when $E_{M} \neq 0$ time-reversal symmetry is broken (a coupled MBS scatterer), breaks time-reversal symmetry as in Eq. ~7, $ s_{12}^{ee} \neq s_{21}^{ee}$. Further, when $E_{M}=0$ (individual MBS), or MBS are absent ($E_{M}=0$, and $\Gamma_{1}= \Gamma_{2}=0$), imply $ s_{12}^{ee} = s_{21}^{ee}$, and TRS is preserved. 
The incoming and outgoing edge modes from MBS scatterer are described in Fig. 1 (c).  When individual MBS are present, i.e. $E_{M} = 0$, in Eq. 7, there is only inter-edge scattering and no intra-edge scattering. Further, in this case, there is no transmission across the STIM junction, only reflection takes place. In the case of coupled MBS, $E_{M} \neq 0$, both intra-edge and inter-edge scattering take place. Further, for coupled MBS, both transmission and reflection take place across the STIM junction.} For the left coupler, the incident edge modes are $I_{1} = (e^{\uparrow}_{o1},h^{\uparrow}_{o1},e^{\uparrow}_{o3},h^{\uparrow}_{o3},e^{\uparrow}_{i5},h^{\uparrow}_{i5})$ (where $e^{\uparrow}_{o1},h^{\uparrow}_{o1}$ are edge modes incoming from the left lead, $e^{\uparrow}_{o3},h^{\uparrow}_{o3}$ are edge modes incoming from the lower arm and $e^{\uparrow}_{i5},h^{\uparrow}_{i5}$ are edge modes incoming from the upper arm) and the corresponding outgoing edge modes are $O_{1} = (e^{\uparrow}_{i1},h^{\uparrow}_{i1},e^{\uparrow}_{i3},h^{\uparrow}_{i3},e^{\uparrow}_{o5},h^{\uparrow}_{o5})$ (shown in Fig. 1(b)). For the right coupler, the incident edge modes are $I_{2}=(e^{\uparrow}_{i2},h^{\uparrow}_{i2},e^{\uparrow}_{i4},h^{\uparrow}_{i4},e^{\uparrow}_{o6},h^{\uparrow}_{o6})$ and the corresponding outgoing edge modes are $O_{2} = (e^{\uparrow}_{o2},h^{\uparrow}_{o2},e^{\uparrow}_{o4},h^{\uparrow}_{o4},e^{\uparrow}_{i6},h^{\uparrow}_{i6})$. The S matrix for the couplers, such that $O_{i} = SI_{i}, i \in \{1,2\}$ is given as \cite{colin},
\begin{equation}
S =
\begin{pmatrix}
-(p+q)I & \sqrt{\epsilon}I & \sqrt{\epsilon}I\\
\sqrt{\epsilon}I & pI & qI\\
\sqrt{\epsilon}I & qI & pI
\end{pmatrix},
\end{equation}
where $p = \dfrac{1}{2}(\sqrt{1 - 2\epsilon} - 1)$ and $q = \dfrac{1}{2}(\sqrt{1 - 2\epsilon} + 1$, $I$ is the $2 \times 2$ identity matrix and $\epsilon$ is a dimensionless parameter which denotes the coupling between the leads and the ring (shown as orange trapeziums) with $\epsilon = 1/2$ for maximum coupling and $\epsilon = 0$ for completely disconnected loop. Edge mode electrons and holes acquire a propagating phase by virtue of traversing the ABI \cite{colin} as follows:\\
In upper arm, left of MBS scatterer,
\begin{equation}
\begin{split}
e^{\uparrow}_{i5} = e^{ik_{e}l_{1}}e^{\dfrac{-i\phi l_{1}}{L}} e^{\uparrow}_{iL},
e^{\uparrow}_{oL} = e^{ik_{e}l_{1}}e^{\dfrac{i\phi l_{1}}{L}} e^{\uparrow}_{o5}, \\
h^{\uparrow}_{i5} = e^{ik_{h}l_{1}}e^{\dfrac{i\phi l_{1}}{L}} h^{\uparrow}_{iL},
h^{\uparrow}_{oL} = e^{ik_{h}l_{1}}e^{\dfrac{-i\phi l_{1}}{L}} h^{\uparrow}_{o5},
\end{split}
\end{equation}
while for upper arm, right of MBS scatterer,
\begin{equation}
\begin{split}
e^{\uparrow}_{o6} = e^{ik_{e}l_{2}}e^{\dfrac{i\phi l_{2}}{L}} e^{\uparrow}_{oR},
e^{\uparrow}_{iR} = e^{ik_{e}l_{2}}e^{\dfrac{-i\phi l_{2}}{L}} e^{\uparrow}_{i6},\\
h^{\uparrow}_{o6} = e^{ik_{h}l_{2}}e^{\dfrac{-i\phi l_{2}}{L}} h^{\uparrow}_{oR},
h^{\uparrow}_{iR} = e^{ik_{h}l_{2}}e^{\dfrac{i\phi l_{2}}{L}} h^{\uparrow}_{i6},
\end{split}
\end{equation}
while for lower arm of ABI,
\begin{equation}
\begin{split}
e^{\uparrow}_{o3} = e^{ik_{e}l_{d}}e^{\dfrac{i\phi l_{d}}{L}} e^{\uparrow}_{o4},
e^{\uparrow}_{i4} = e^{ik_{d}l_{d}}e^{\dfrac{-i\phi l_{d}}{L}} e^{\uparrow}_{i3},\\
h^{\uparrow}_{o3} = e^{ik_{h}l_{d}}e^{\dfrac{-i\phi l_{d}}{L}} h^{\uparrow}_{o4},
h^{\uparrow}_{i4} = e^{ik_{h}l_{d}}e^{\dfrac{i\phi l_{d}}{L}} h^{\uparrow}_{i3},
\end{split}
\end{equation}
\begin{figure}
\center
\includegraphics[width=.5\textwidth]{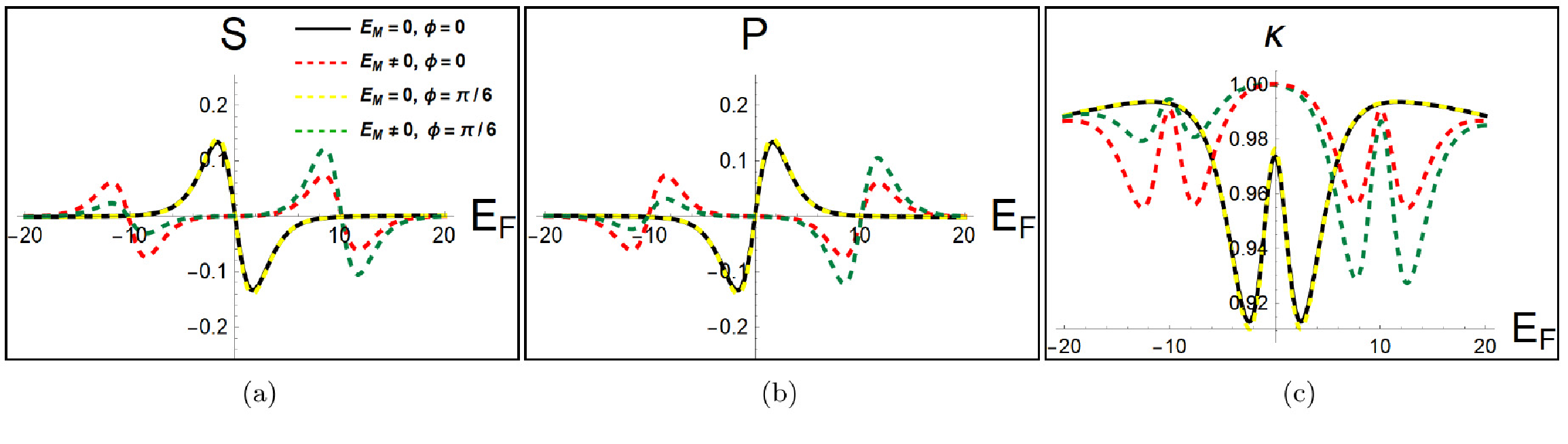}
\caption{Seebeck coefficient $S$ in units of $k_{B}/eT$, (b) Peltier coefficient $P$ in units of $k_{B}/e$ and (c) Thermal Conductance $\kappa$ in units of $\kappa_{0}$ vs. Fermi energy $E_{F}$ with $\phi = 0$ and $\phi = (\pi/6)$ with temperature $T = 115mK$, plotted for $E_{M} = 0$ (individual MBS) and $E_{M} = 10 \mu eV$ (coupled MBS).}
\end{figure}
with $k_{e} = (E + E_{f})/\hbar v_{F}$ and $k_{h} = (E - E_{f})/\hbar v_{F}$. $L$ is the total length of ring, while $l_{1}$ and $l_{2}$ are lengths of left and right part of upper branch respectively, and $l_{u}$, $l_{d}$ are lengths of the upper and lower branches. $\phi$ is the modified flux parameter given as $\phi = \Phi/\Phi_{0}$, where $\Phi_{0}$ is the flux quantum $hc/e$.\\
A similar ABI can distinguish between coupled and uncoupled MBS using the electrical conductance \cite{colin}. In the next section, we will look at the behavior of various thermoelectric coefficients in the presence and absence of MBS to detect MBS.
\begin{figure}
\center
\includegraphics[width=.5\textwidth]{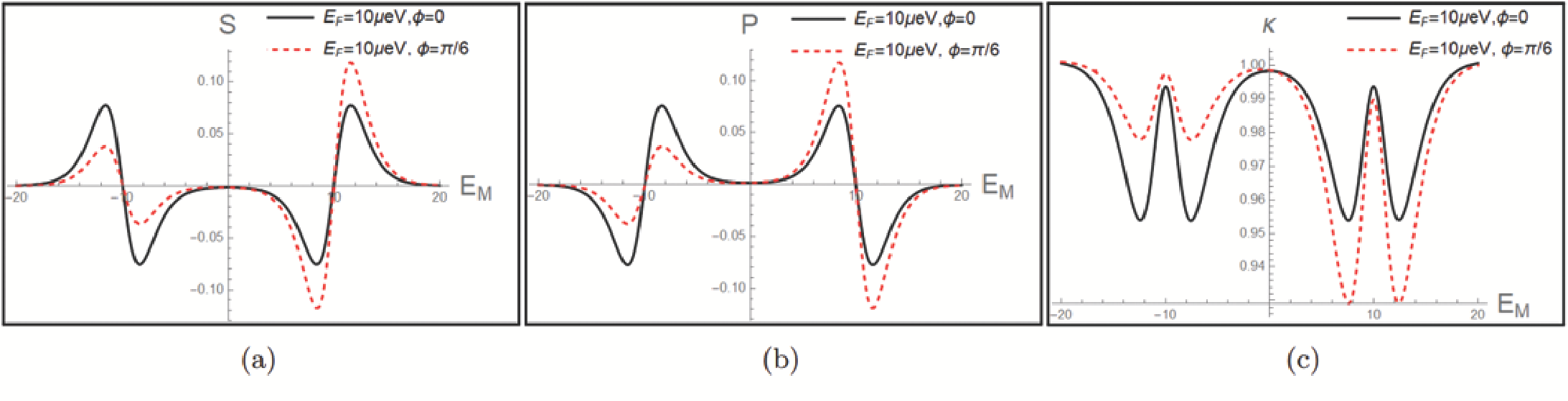}
\caption{Seebeck coefficient $S$ in units of $k_{B}/eT$, (b) Peltier coefficient $P$ in units of $k_{B}/e$ and (c) Thermal Conductance $\kappa$ in units of $\kappa_{0}$ vs. coupling $E_{M} $between MBS for $\phi = 0$ and $\phi = (\pi/6)$ with temperature $T = 115mK$.}
\end{figure}
\section{Signatures of Majorana bound states in thermoelectric coefficients}
\subsection{Thermoelectric transport in Aharonov-Bohm interferometer with and without MBS}
We determine the total transmission probability for the Aharonov-Bohm ring with MBS using Eqs. (7-11) to solve the scattering problem. Substituting the transmission probability in Eq. (1), we calculate the Onsager coefficients and various other thermoelectric coefficients from Eqs. (3, 4), plot them versus Aharonov-Bohm flux and the Fermi energy. In our calculations, we use the following values for the parameters, $\epsilon = 0.5$, $\Gamma_{1} = \Gamma_{2} = 1 \mu eV$, $l_{u} = l_{d} = L/2 = 5 \mu m$ and the Aharonov-Bohm flux in units of the flux quanta $hc/e$. A characteristic experimental value for coupling strength between Majorana bound states $E_{M}$ is of the order of $10 \mu eV$ \cite{fuKaneMBS, nilssonMBS}.
\begin{table*}
\begin{center}
{\begin{tabular}{|l|l|l|l|}
\hline
\multicolumn{1}{|c|}{Thermoelectric Parameter} & \multicolumn{1}{c|}{MBS} & \multicolumn{1}{c|}{Magnetic Field($\phi$) asymmetry} & \multicolumn{1}{c|}{Fermi Energy($E_{F}$) asymmetry}\\ \hline
\multicolumn{1}{|c|}{} & \multicolumn{1}{c|}{Absent} & \multicolumn{1}{c|}{$S(\phi) = S(-\phi) \approx 0$} & \multicolumn{1}{c|}{$S(E_F) = -S(-E_F) \approx 0$} \\ \cline{2-4}
\multicolumn{1}{|c|}{{Seebeck Coefficient (S)}} & Individual, $E_{M} = 0$ & \multicolumn{1}{c|}{$S(\phi) = S(-\phi)\neq 0$} & \multicolumn{1}{c|}{$S(E_F) = -S(-E_F)\neq 0$} \\ \cline{2-4}
\multicolumn{1}{|c|}{} & Coupled, $E_{M}\neq 0$ & \multicolumn{1}{c|}{$S(\phi) \neq S(-\phi)$} & \multicolumn{1}{c|}{$S(E_F) \neq -S(-E_F)$} \\ \hline
\multicolumn{1}{|c|}{} & \multicolumn{1}{c|}{Absent} & \multicolumn{1}{c|}{$P(\phi) = P(-\phi) \approx 0$} & \multicolumn{1}{c|}{$P(E_F)=-P(-E_F) \approx 0$} \\ \cline{2-4}
\multicolumn{1}{|c|}{{Peltier Coefficient (P)}} & \multicolumn{1}{c|}{Individual, $E_{M}=0$} & \multicolumn{1}{c|}{$P(\phi) = P(-\phi)\neq 0$} & \multicolumn{1}{c|}{$P(E_F)=-P(-E_F)\neq 0$} \\ \cline{2-4}
\multicolumn{1}{|c|}{} & Coupled, $E_{M}\neq 0$ & \multicolumn{1}{c|}{$P(\phi) \neq P(-\phi)$} & \multicolumn{1}{c|}{$P(E_F)\neq-P(-E_F)$} \\ \hline
\multicolumn{1}{|c|}{}&\multicolumn{1}{c|}{Absent}&\multicolumn{1}{c|}{$\kappa(\phi) = \kappa(-\phi) \approx \kappa_{0}$}&\multicolumn{1}{c|}{$\kappa(E_F)=\kappa(-E_F) \approx \kappa_{0}$} \\ \cline{2-4}
\multicolumn{1}{|c|}{{Thermal conductance $(\kappa)$}}&\multicolumn{1}{c|}{Individual, $E_{M}=0$}&\multicolumn{1}{c|}{$\kappa(\phi) = \kappa(-\phi)\neq \kappa_{0} $}&\multicolumn{1}{c|}{$\kappa(E_F)=\kappa(-E_F)\neq \kappa_{0}$} \\ \cline{2-4}
\multicolumn{1}{|c|}{} & Coupled, $E_{M}\neq 0$ & \multicolumn{1}{c|}{$\kappa(\phi) \neq \kappa(-\phi)$} & \multicolumn{1}{c|}{$\kappa(E_F) \neq \kappa(-E_F)$} \\ \hline
\multicolumn{1}{|c|}{}&\multicolumn{1}{c|}{Absent}&\multicolumn{1}{c|}{Preserved}&\multicolumn{1}{c|}{Preserved} \\ \cline{2-4}
\multicolumn{1}{|c|}{{WF Law $(W)$}}&\multicolumn{1}{c|}{Individual, $E_{M}=0$}&\multicolumn{1}{c|}{Violated, $W(\phi) = W(-\phi)$}&\multicolumn{1}{c|}{Violated, $W(E_F)=W(-E_F)$} \\ \cline{2-4}
\multicolumn{1}{|c|}{} & Coupled, $E_{M}\neq 0$ & \multicolumn{1}{c|}{Violated, $W(\phi) \neq W(-\phi)$} & \multicolumn{1}{c|}{Violated, $W(E_F) \neq W(-E_F)$} \\ \hline
\end{tabular}}
\end{center}
\caption{Probing MBS. The magnetic field asymmetry is shown at $E_{F} = 3 \mu eV$ for Seebeck coefficient, Peltier coefficient, and thermal conductance, and at $E_{F} = 10 \mu eV$ for WF law. The Fermi energy asymmetry is shown at $\phi = (\pi/6)$}
\end{table*}
\begin{figure}
\center
\includegraphics[width=0.5\textwidth]{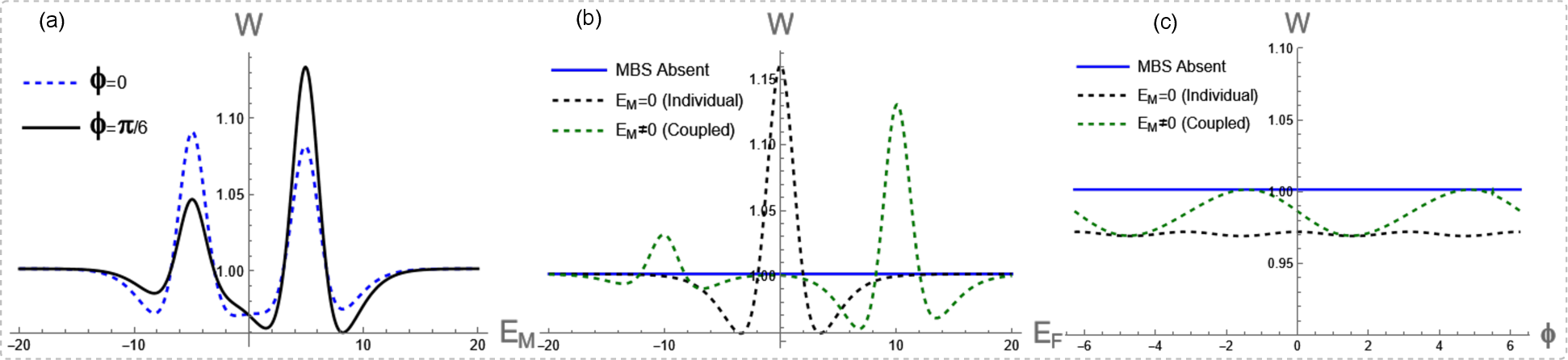}
\caption{The Lorenz ratio $W$ (a) vs. $E_{M}$ for different values of flux $\phi$, (b) vs. $E_F$ for different values of $E_{M}$, and (c) vs. $\phi$ for different values of $E_M$, with $T = 115mK$. For $E_{M} \neq 0$ (coupled MBS) case we take $E_M = 10\mu eV$ in (b) and (c), while $E_F = 5 \mu eV$ in (a) and (c) and $\phi=\pi/6$ in (b).}
\end{figure}
In Fig. 2, we have plotted various thermoelectric coefficients against Aharonov-Bohm flux $\phi$ for $E_{F} =3\mu eV$, and for different values of $E_{M}$. The value of Fermi energy is $3\mu eV$ as all three thermoelectric coefficients have comparable magnitude at $E_{F} = 3\mu eV$ for all cases. This paper considers the following three cases: MBS are absent, MBS are present and uncoupled, and MBS are present and coupled. MBS occurs when the superconducting-ferromagnetic junction is present in the system (STIM junction). When the STIM junction is absent, MBS does not occur, and the matrix in Eq. (7) becomes an identity matrix. When the STIM junction is absent, the Seebeck and Peltier coefficients vanish as Andreev-reflection, time-reversal symmetry (TRS) remains unbroken, and the thermal conductance is constant. When the STIM junction is added to the system, MBS occurs. We can see from Fig. 2 that when MBS are uncoupled ($E_{M} = 0$), the Seebeck and Peltier coefficients behave as periodic functions but, more importantly, as symmetric functions of Aharonov-Bohm flux $(\phi)$ and the thermal conductance also behaves as a symmetric function of $\phi$. For coupled MBS ($E_{M} \neq 0$), the Seebeck and Peltier coefficients and thermal conductance are asymmetric to $\phi$. Further, we find the distinction sharper for higher values of $E_{m}$.\\
In Fig. 3, we plotted the thermoelectric coefficients versus Fermi energy $E_{F}$ for coupled and uncoupled MBS at different flux values $\phi$. We can see that for a finite value of $\phi$, the thermal conductance behaves as a symmetric function of Fermi energy $E_{F}$, while Seebeck and Peltier's coefficients act as anti-symmetric functions of Fermi energy when MBS are uncoupled ($E_{M} = 0$). When MBS are coupled ($E_{M} \neq 0$), the thermoelectric coefficients are asymmetric to the Fermi energy. For $\phi = 0$, however, the symmetry in thermal conductance and anti-symmetry in Seebeck and Peltier coefficients to Fermi energy is seen regardless of whether MBS are coupled. We see that the Seebeck and Peltier coefficients change sign, and the thermal conductance displays peaks at $E_{F} = \pm E_{M}$ when MBS are coupled. This clearly demonstrates the splitting of the Majorana zero mode to $+E_{M}$, and $-E_{M}$ when MBS are coupled \cite{nilssonMBS}.\\
In Fig. 4, we plot the Seebeck, Peltier and thermal conductance versus the coupling of the MBS $E_{M}$. We see from Fig. 4 that the Seebeck coefficient, Peltier coefficient, and the thermal conductance all behave as symmetric functions of the coupling $E_{M}$ when flux is absent. However, when a flux is present in the setup, time-reversal symmetry is broken and the thermoelectric coefficients behave asymmetrically  with respect to sign reversal of $E_{M}$.
From Figs. 2 (a) and (b), we see that coupled MBS cause Seebeck and Peltier coefficients to change signs, e.g., for negative charge carriers ($E_{F} > 0$), we expect the Seebeck coefficient to be negative. The Peltier coefficient is positive and vice-versa. However, when $E_{M} \neq 0$, we see that the signs of the Seebeck and Peltier coefficient are different from our expectations. We can see this more clearly in Figs. 3 (a) and (b) where we see that in presence of MBS, the signs are opposite to our expectation for $-E_{M} < E_{F} < E_{M}$. This behavior is a signature of MBS.\\
Finally, when MBS are absent, the matrix in Eq. (7) becomes an identity. In this case, we have not plotted the coefficients as the Seebeck and Peltier coefficients vanish, and thermal conductance remains at a near-constant value of $\kappa_{0} = \pi^{2} k_{B}^{2} T/3h$ for the entire range of $\phi$ in Fig. 2 and the whole range of $E_{F}$ in Fig. 3.
\subsection{Identifying MBS}
As discussed in the previous subsection, we see that for uncoupled MBS, the Seebeck and Peltier coefficients and the thermal conductance behave as a symmetric function of $\phi$. This symmetry is absent when MBS are coupled. The absence of symmetry is primarily caused by the breaking of TRS by AB flux \cite{BUTTIKER1983365} in the presence of a coupled MBS scatterer, e.g. in Eq. (7), $y(E) \neq y(-E)$ when $E_{M} \neq 0$. Due to the breakdown of TRS, electrons and holes scattered acquire different phases while traversing the TI in opposite directions. Thus, asymmetric plots for thermoelectric coefficients result when an AB flux and coupled MBS are present. Similarly, the thermoelectric coefficients are asymmetric to the reversal in Fermi energy only when both coupled MBS, and an AB flux is present. Suppose MBS are uncoupled, and flux is absent. In that case, the Seebeck and Peltier coefficients are anti-symmetric functions of Fermi energy, and the thermal conductance is a symmetric function of Fermi energy. We summarize these results in Table 1. We can see that both Aharonov-Bohm flux and Fermi energy are valuable measures for probing MBS. When MBS are absent in the system, both TRS and electron-hole symmetry are preserved, and Seebeck and Peltier coefficients vanish while the thermal conductance remains nearly constant.
\section{Wiedemann-Franz law}
Wiedemann–Franz (WF) law states that the ratio of the thermal conductance ($\kappa$) to the electric conductance ($\sigma$) is proportional to the temperature \cite{weideman} of the system. Thus,
\begin{equation}
\kappa/\sigma = W, \mbox{where $W$ is the Lorenz ratio.}
\end{equation}
 When WF law is preserved, $W = \kappa_{0}/G_{0}$ is a constant. Wiedemann-Franz law in presence and absence of coupled MBS has been studied before \cite{ramosweideman, ricco2018tuning}. For the systems studied in \cite{ricco2018tuning} and one of the systems studied in \cite{ramosweideman}, the violation in Wiedemann-Franz law occurs only when MBS are present and coupled. The transmission probability of a different model discussed in \cite{ramosweideman} is similar in nature to that of our model. We will now look at the violation of WF law in our setup. {An important point regarding Fig. 5 (a), is that we plot the Lorenz ratio $W$ vs. coupling $E_{M}$ in the presence as well as in the absence of magnetic flux. We see that violations in WF law occur in the presence of MBS. When MBS are coupled, we see peaks in $W$ at $E_{F} = \pm E_{M}$. Again, this is a consequence of the splitting of the Majorana bound state to $+E_{M}$, and $-E_{M}$ states. Similar violations of WF law have been reported before in systems hosting MBS \cite{ramosweideman}. This distinct behavior of WF law in all three cases (absent, individual, and coupled MBS) has not been reported before.\\
In Fig. 5(b), $W$ is plotted as function of the Fermi energy. We see violations for both coupled and individual MBS. In Fig. 5 (c), we plot $W$ versus flux $\phi$. Once again, we see that WF law is preserved when MBS are absent and violated when MBS are present. The violation again occurs at $E_{F} \approx \pm E_{M}$. Further, similar to the other thermoelectric coefficients, we see that the Lorenz ratio $W$ is symmetric to flux ($\phi$) reversal when MBS are uncoupled and asymmetric when MBS are coupled. (c). We observe that when MBS are absent, there is no violation. However, when MBS are present, violations of WF law are observed regardless of their coupling. Further, we also observe that $W$ behaves similarly as the other thermoelectric coefficients. $W$ is symmetric to the reversal in Fermi energy when MBS are uncoupled and asymmetric when MBS are coupled. }
\section{Conclusion} In this paper, we have presented a method to probe the existence and nature of MBS using various thermoelectric coefficients and through the violation of WF law. In our proposed setup, we see that the presence of both an AB flux and coupled MBS results in breaking TRS. We plot the thermoelectric coefficients: Seebeck coefficient, Peltier coefficient, and the thermal conductance against flux $\phi$. Seebeck and Peltier coefficients and thermal conductance behave as symmetric functions of $\phi$ when MBS are uncoupled. When MBS are coupled, the coefficients behave asymmetrically to $\phi$. Similarly, in the absence of flux $\phi$, Seebeck and Peltier coefficients behave anti-symmetrically while thermal conductance behaves symmetrically to Fermi energy reversal regardless of the coupling of MBS. However, when the flux is present, the symmetry in thermal conductance and the anti-symmetry in Seebeck and Peltier coefficients are observed when MBS are uncoupled. When MBS are coupled, the Seebeck and Peltier coefficients and the thermal conductance are asymmetric functions of Fermi energy. This distinction allows us to use Aharonov-Bohm flux and Fermi energy to probe the existence and nature of MBS. Another parameter we have studied in this work is the violation of the Wiedemann-Franz law. We see that the violation occurs only in the presence of MBS regardless of their coupling. The violation occurs roughly at $E_{F} \approx \pm E_{M}$. We also see that $W$ is symmetric to $E_{F}$ when MBS are uncoupled, flux is absent, and asymmetric when MBS are coupled, and a non-zero flux is present.
Similarly, when $W$ is plotted vs. $\phi$, we see that when MBS are uncoupled, $W$ is symmetric to $\phi$ and asymmetric when MBS are coupled. Finally, when MBS are absent, the Seebeck and Peltier coefficients vanish, and thermal conductance remains at a near-constant value of $\kappa_{0}$, while the Wiedemann-Franz law is preserved. The proposed setup may be realized in a HgTe quantum well with a field of 0.03T inside the ferromagnets \cite{nilssonMBS}. {The length of the TI-AB ring is much greater than the length of the ferromagnetic/superconducting/ferromagnetic (FM/SC/FM) trilayer, further, the ferromagnetic layers should be very thin compared to the superconducting slip so that, effectively, the phase accumulated by the electrons/holes due to the ferromagnetic layer is ignored.} In the future, we plan to look at the performance of similar systems hosting MBS as potential thermoelectric heat engines and refrigerators.
\acknowledgements
The grant which supported this work: Josephson junctions with strained Dirac materials and their application in quantum information processing, Science \& Engineering Research Board (SERB) DST, Govt. of India, Grant No. CRG/20l9/006258.
\section{Data Availability Statement} The data supporting this study’s findings are available within the article.

\end{document}